\documentstyle[pra,preprint,aps]{revtex}

\newcommand{\rme}{{\rm e}}
\newcommand{\rmi}{{\rm i}}
\newcommand{\rmd}{{\rm d}}

\begin{document}

\draft

\title{Classical and quantum dynamics of a spin-$\frac{1}{2}$}

\author{Adrian Alscher and Hermann Grabert}

\address{Fakult\"at f\"ur Physik, Albert-Ludwigs-Universit\"at,\\
Hermann-Herder-Strasse 3, D-79104 Freiburg, Germany}
\date{\today}
\maketitle 
\tighten
\begin{abstract}
We reply to a comment on `Semiclassical dynamics of a spin-$\frac{1}{2}$
in an arbitrary magnetic field'.
\end{abstract}

\pacs{03.65.Sq, 67.57.Lm, 31.15.Kb}

\narrowtext                              

In a recent Comment \cite{kochetov1} Kochetov argues that our 
results \cite{alscher} on the coherent state path integral for a
spin-$\frac{1}{2}$ in an arbitrary magnetic field are based on a
`classical spin action inconsistent with the necessary boundary
conditions'. 
Contrary from what is insinuated by the Comment our article
is not concerned with the {\it quantization} of a classical spin but
solely with the {\it representation} of a {\it quantum}
spin-$\frac{1}{2}$ in terms of a spin coherent state path
integral. Hence, the Comment by Kochetov arguing primarily on a
classical level is only vaguely relevant to our work and
chiefly reconsiders the author's earlier work
\cite{kochetov2,kochetov3} in the light of results in
\cite{alscher}. Indeed, the analysis in \cite{alscher} allows for some
conclusions of relevance to Kochetov's work as discussed below. 

Before addressing the Comment more specifically let us briefly
reformulate the approach in \cite{alscher}. 
We  start with the two-dimensional Hilbert space, 
represented in the basis of spin coherent states 
$\bigl|\Psi_g\bigr>  = {\cal D}^{1/2}(g)\bigl|\uparrow\bigr>$, 
where $g\in SU(2)$ \cite{perelomov}.
Since $\bigl|\Psi_h\bigr>  = \exp(\rmi \alpha)\bigl|\uparrow\bigr>$ 
for a  $h$ in the maximal torus, 
$\bigl|\Psi_g\bigr>$ and $\bigl|\Psi_{g'}\bigr>$ describe the same 
physical state if there exists a $h \in U(1)$ such that $g'=gh$. 
Therefore the group $SU(2)$ can be viewed as
fibre bundle over the base manifold $SU(2)/U(1) \equiv S^2$ with fibre $U(1)$
\cite{balachandran} and the space of distinct spin coherent states is
canonically isomorphic to these left cosets. 
Parametrizing any $g\in SU(2)$ with Euler angles $(\vartheta,\varphi,\chi)$, we get
\begin{equation}
\bigl|\Omega\bigr>  =\rme^{-\frac{\rmi}{2}\chi}\rme^{-\rmi\varphi S_{z}}
\rme^{-\rmi \vartheta S_{y}}\bigl|\uparrow\bigr>.
\label{eq20} 
\end{equation}
Here, the first factor on the righthand side is just a phase factor
and the rest determines the physical state. 
These states are not orthogonal but form an overcomplete basis in the
Hilbert space. The overlap is readily evaluated and the identity may 
be represented as
\begin{equation}
I =\frac{1}{2\pi}\int \sin(\vartheta)\rmd\vartheta \rmd\varphi 
\left|\Omega\right>\left<\Omega\right|.
\label{eq22b}  
\end{equation}
Employing a Trotter decomposition, the propagator may be written as 
\begin{eqnarray}
\bigl<\Omega''\bigr|U(t)\bigl|\Omega'\bigr>&=&
\lim_{\epsilon\to 0} N 
\int\prod_{k=1}^{n} \sqrt{\det{\omega_{ij}}}\rmd\vartheta_k \rmd\varphi_k
\nonumber\\ 
&&
\times
\exp\left\{\sum_{k=0}^{n}
\left[
\log{\bigl<\Omega_{k+1}\bigr.\bigl|\Omega_{k}\bigr>} +
\frac{\rmi}{2}(\chi_{k+1}-\chi_k) -
\rmi\epsilon\frac{
\bigl<\Omega_{k+1}\bigr|H(k\epsilon)\bigl|\Omega_{k}\bigr>}
{\bigl<\Omega_{k+1}\bigr.\bigl|\Omega_{k}\bigr>} 
\right]\right\}.
\label{eq23} 
\end{eqnarray}
where $\epsilon=t/n$, $(\Omega_{0},\chi_{0})=(\Omega',\chi')$, 
$(\Omega_{n+1},\chi_{n+1})=(\Omega'',\chi'')$.  We are allowed to pass
to the continuum limit, if the paths stay continuous  
for $\epsilon\rightarrow 0$ which is not guaranteed if no Wiener measure
occurs. Therefore care must be taken in calculating the path
integral \cite{solari}-\cite{shibata}.
To ensure an integration over continuous Brownian
motion paths, we introduce a regularization by the spherical Wiener measure
and are then allowed to write
\begin{eqnarray}
\bigl<\Omega''\bigr|U(t)\bigl|\Omega'\bigr>&=&\lim_{\nu\to\infty} 
N \int \prod_{s=0}^{t} \sqrt{\det{\omega_{ij}}} \rmd\vartheta(s) \rmd\varphi(s) 
\nonumber\\
&&\times
 \exp{ \biggl\{ \rmi\int_{0}^{t} \rmd s 
\Bigl[ 
\frac{\rmi}{\nu}(g_{\vartheta\vartheta}{\dot\vartheta}^2
+g_{\varphi\varphi}{\dot\varphi}^2) 
+\theta_{\vartheta}\dot\vartheta+\theta_{\varphi}\dot\varphi
-H(\vartheta,\varphi,s)\Bigr]\biggr\} }.
\label{eq29}
\end{eqnarray}
Here, $N =\lim_{n\to\infty}\prod_{k=1}^{n}\frac{1}{\pi}$ is a normalization factor, 
$g=\frac{1}{4}(\rmd \vartheta\otimes\rmd \vartheta +
\sin(\vartheta)^2 \rmd \varphi\otimes\rmd \varphi)$ the
metrical tensor, 
$\omega=\frac{1}{2}\sin(\vartheta)\rmd \vartheta \wedge \rmd \varphi$ 
the symplectic two-form of $SU(2)/U(1)$ \cite{provost} 
and $\theta=\frac{1}{2}(\cos(\vartheta)\rmd\varphi +\rmd \chi)$ its
corresponding symplectic potential ($\omega=-\rmd\theta$).

Choosing in every left coset one special representant, i.e.
fixing $\chi$ for every coherent state, one defines a section of the
$SU(2)$ bundle. In particular, the choice $\chi=0$ was adopted in
\cite{alscher}. It is important to note that  {\it once $\chi$ has been
fixed the symplectic potential is  fixed as well} and
manipulations of the form suggested by Kochetov \cite{kochetov1} in
equation (3) are no longer allowed. The very same reasoning applies in
the parametrization used by Kochetov. Within the Gaussian decomposition
\cite{perelomov} of the elements of $SU(2)$ by
$g=z_{-}hz_{+}$ for $z_{-}\in Z_{-}$, $h\in U(1)$, $z_{+}\in Z_{+}$  or equivalently
$g=z_{-}b_{+}$ with $b_{+}\in B_{+}$, we recognize that 
${\cal D}^{1/2}(g)\bigl|\uparrow\bigr>={\cal D}^{1/2}(z_{-}h)\bigl|\uparrow\bigr>$.
Parametrizing $z_{-}$ by the complex number $\zeta$, the space 
of distinct spin coherent states is now 
isomorphic to elements in $SL(2,{\cal C })/B_{+} \cong
SU(2)/U(1)$. If the isomorphismus is defined explicitly
by the spherical projection from the south pole of the sphere onto the complex
plane, one has
$\zeta=\tan\left(\frac{\vartheta}{2}\right)\rme^{\rmi\varphi}$, and 
makes use of a {\it different} section of the $SU(2)$ bundle by
setting $\chi=-\varphi$. Therefore a corresponding phase factor appears
\begin{equation}
\label{eq33a}
\bigl|\zeta\bigr>  =\frac{1}{\sqrt{1+|\zeta|^2}}\rme^{\zeta S_{-}}
\bigl|\uparrow\bigr>=\rme^{\frac{\rmi}{2}\varphi}\bigl|\Omega\bigr>.
\end{equation}
Again with the choice $\chi=-\varphi$ there is no room for additional
manipulations of the form (3) in \cite{kochetov1}. While Kochetov's 
theory starts from a classical spin and employs geometric
quantization to obtain a quantum propagator after an {\it ad
hoc} modification of the symplectic potential, no such ambiguities
arise if the representation of the quantum propagator in terms of a
path integral is considered.

A main point in the critique by Kochetov \cite{kochetov1} is the claim
that the approach in \cite{alscher} disagrees with boundary
conditions  in the classical limit. Since the physical states form a 
symplectic two-dimensional differential manifold with the closed 
two-form $\omega$, the classical dynamics is determined by the 
Hamiltonian vector field $\omega(X_H,\cdot)=\rmd H$  which leads 
immediately to the classical equations of motion (23) in
\cite{alscher}. Note that in general  there is {\it no classical path}
connecting arbitrary but real boundary conditions 
$\bar\Omega(0)=\bar\Omega'$ and $\bar\Omega(t)=\bar\Omega''$. 
This is known as the `overspecification problem'. Kochetov modifies 
the action to allow always for a `classical' path, which is usually 
complex. If one applied the same rules to a simple harmonic oscillator
there would be a `classical' path connecting any initial phase space
point $(q',p')$ with any endpoint $(q'',p'')$. This is clearly not
what is usually meant by classical. Hence, the overspecification problem
should {\it not} be removed in the classical limit. Yet, in the
quantum problem, there is indeed a semiclassical path for any pair of 
real boundary conditions [see equations (23) {\it and}\, (24),(25) in 
\cite{alscher}].

Finally, Kochetov believes that the exactness of the semiclassical 
propagator is `obvious' and `self-evident'. Replacing Kochetov's 
qualitative arguments by a more accurate treatment \cite{niemi,szabo}, one  
finds the  necessary condition $\theta({X_{H})}=H$ that
$\theta$ is $SU(2)$ invariant and the stationary phase approximation
becomes exact. For a Hamilton operator 
which is a linear combination of all three generators of the $su(2)$
algebra  we get 
three conditions which cannot be satisfied generally by 
fixing the phase $\chi$ appropriately. Therefore there are no $SU(2)$
invariant potentials on $S^2$ and although  $SU(2)$ is the group of  
isometric canonical transformations on the two-sphere \cite{kobayashi},
it does not preserve the sections. Hence, for magnetic fields of
arbitrary time-dependence, the exactness of dominant stationary phase
approximation (DSPA) is not ``self-evident'', and prior to our work
\cite{alscher} it was rather expected that the DSPA does not provide a
correct result \cite{klauder}.

\end{document}